\documentclass{JCIN22}
\usepackage{lettrine}
\usepackage{setspace}
\makeatletter
\def\@upcite#1#2{\textsuperscript{[{#1\if@tempswa , #2\fi}]}}
\makeatother

\begin{document}

\ArticleType{Research paper}
\Year{2022}

\title{LLM4CP: Adapting Large Language Models for Channel Prediction}

\author{Boxun Liu}
\author{Xuanyu Liu}
\author{Shijian Gao}
\author{Xiang Cheng}
\author{Liuqing Yang}

\maketitle

{\bf\textit{Abstract---}
Channel prediction is an effective approach for reducing the feedback or estimation overhead in massive multi-input multi-output (m-MIMO) systems. 
However, existing channel prediction methods lack precision due to model mismatch errors or network generalization issues.
Large language models (LLMs) have demonstrated powerful modeling and generalization abilities, and have been successfully applied to cross-modal tasks, including the time series analysis.
Leveraging the expressive power of LLMs, we propose a pre-trained LLM-empowered channel prediction method (LLM4CP) to predict the future downlink channel state information (CSI) sequence based on the historical uplink CSI sequence.
We fine-tune the network while freezing most of the parameters of the pre-trained LLM for better cross-modality knowledge transfer.
To bridge the gap between the channel data and the feature space of the LLM, preprocessor, embedding, and output modules are specifically tailored by taking into account unique channel characteristics.
Simulations validate that the proposed method achieves SOTA prediction performance on full-sample, few-shot, and generalization tests with low training and inference costs.

\textit{Keywords---}
Channel prediction, massive multi-input multi-output (m-MIMO), large language models (LLMs), fine-tuning, time-series
}

\section{Introduction}
\lettrine[lines=2]{M}{}assive multi-input multi-output (MIMO) technology is regarded as a core technology in the fifth-generation (5G) and beyond 5G mobile communication systems\upcite{cheng2023intelligent} for enhancing the spectral efficiency (SE).
Accurate channel state information (CSI) plays a fundamental role in facilitating m-MIMO related design, such as transceiver optimization\upcite{zhang2023integrated}, adaptive modulation\upcite{chung2001degrees}, resource allocation\upcite{sadr2009radio}, and so on.
Typically, CSI is acquired through channel estimation\upcite{gao2020estimating,ma20003optimal} whose updating frequency is dictated by the channel coherence time.
For mobility scenarios involving high-velocity users, the shortened channel coherence time significantly increases the overhead of channel estimation, thereby leading to an appreciable reduction in the system SE.
Additionally, in frequency-division duplex (FDD) systems where channel reciprocity does not hold, the base station (BS) can only obtain downlink CSI through user feedback, resulting in increased overhead and delay.

Channel prediction\upcite{rottenberg2020performance,choi2020experimental,arnold2019towards} is a promising technology to reduce the CSI acquisition overhead, which predicts future CSI based on historical CSI data. 
The historical CSI and the predicted CSI can either be in the same or different frequency bands, corresponding to frequency-division duplex (TDD) and FDD modes, respectively.
For instance, in FDD systems, the downlink CSI can be inferred from previous uplink CSI, thereby avoiding the need for channel estimation and feedback.
Existing studies on channel prediction can be categorized into three types, i.e., model-based methods, deep learning-based methods, and hybrid (physics-informed deep learning-based) methods.

For model-based methods, several parametric models have been investigated for temporal channel prediction, including the autoregressive (AR) model\upcite{truong2013effects}, the sum-of-sinusoids model\upcite{wong2005joint}, and the polynomial extrapolation model\upcite{shen2003short}.
In Ref. \cite{yin2020addressing}, a Prony-based angular-delay domain (PAD) channel prediction algorithm was proposed by exploiting the high resolution of multipath angle and delay in massive MIMO-OFDM systems.
Additionally, a joint angle-delay-Doppler (JADD) CSI acquisition framework was designed\upcite{qin2022partial} for FDD systems, utilizing the partial reciprocity between uplink and downlink channels.
Nevertheless, the effectiveness of model-based approaches is heavily dependent on the accuracy of the theoretical model, which can be challenging to fit into the complex multipath characteristics of the practical channel. 

Deep learning has demonstrated its powerful capabilities in automatically adapting to data distribution without prior assumptions.
Recently, several classical neural networks\upcite{kim2020massive,jiang2019neural,jiang2022accurate,jiang2020deep,safari2019deep,zhang2023adversarial} have been applied for channel prediction tasks.
In Ref. \cite{kim2020massive}, a multi-layer perceptron (MLP)-based channel prediction method demonstrates comparable performance to vector Kalman filter (VKF)-based channel predictor.
To better learn temporal variations, recurrent neural network\upcite{jiang2019neural} and long short-term memory (LSTM)\upcite{jiang2020deep} are applied to channel prediction.
In addition, a transformer-based parallel channel prediction scheme\upcite{jiang2022accurate} was proposed to avoid error propagation in the sequential CSI prediction process. 
In Ref. \cite{safari2019deep,zhang2023adversarial}, convolutional neural networks (CNN) and generative adversarial networks (GANs) are utilized for downlink CSI prediction by treating the prediction process as image processing.
However, due to the lack of consideration for the unique structure of the channel, the above methods struggle to handle complex channel prediction tasks and exhibit high complexity.

Therefore, a few physics-informed deep learning-based\upcite{gao2021model,fan2024deep} works have considered the unique characteristics of CSI, referred to as hybrid methods\upcite{burghal2023enhanced,liu2022spatio}.
For instance, a 3-Dimensional (3D) complex CNN-based predictor\upcite{burghal2023enhanced} is utilized to capture temporal and spatial correlations based on angle-delay domain representation.
In Ref. \cite{liu2022spatio}, a ConvLSTM-based spatio-temporal neural network (STNN) is proposed to parallel process high-dimensional spatial-temporal CSI.
Nevertheless, the scalability of hybrid methods is relatively poor, requiring sufficient understanding of the channel structure.

Despite significant progress achieved by deep learning-empowered methods, there are still some shortcomings limiting their application in practical scenarios.
First, the predictive capability is constrained by the size of networks.
Existing methods struggle to accurately model complex spatial, temporal, and frequency relationships, especially for high-velocity scenarios and FDD systems.
Secondly, in contrast to model-based approaches, deep learning-based methods exhibit poor generalization ability, requiring retraining when the CSI distribution changes. 
Although a few studies aim to improve generalization ability by meta-learning\upcite{kim2023massive} or hypernetwork\upcite{liu2024hypernetwork}, the additional adaptation stage or the hypernetwork branch increases the operational complexity.
In summary, existing deep learning-empowered prediction models struggle to meet the requirements for high generalization performance and accurate prediction capabilities.

Large language models (LLMs) have achieved tremendous success in the field of natural language processing (NLP) and have led to a new paradigm, namely, fine-tuning models pre-trained on large-scale datasets for downstream tasks with few or zero labels.
This provides a promising solution to address the shortcomings of existing channel prediction schemes.
However, previous downstream tasks were limited to the field of NLP.
Recently, several studies\upcite{liang2024foundation,su2024large,zhou2024one,jin2023time,ren2024tpllm} have provided initial evidence of the powerful cross-modal transfer capabilities of pre-trained LLMs.
For instance, Ref. \cite{zhou2024one} fine-tunes frozen pre-trained LLM on time series datasets and achieves state-of-the-art (SOTA) performance on main time series analysis tasks.
In Ref. \cite{ren2024tpllm}, a traffic prediction framework based on pre-trained LLM (TPLLM) is proposed with the low-rank adaptation (LoRA) fine-tuning approach.
Still, existing cross-modal fine-tuning works mainly focused on temporal or spatio-temporal series prediction rather than channel prediction tasks.
Unlike time series forecasting tasks, there are certain difficulties in adapting LLM for channel prediction\upcite{zhou2024large}.
First, CSI is the high-dimensional structural data with the multi-path channel model rather than simple one-dimensional data, which increases the complexity of processing.
Moreover, there is a huge domain gap between CSI and the natural language. 
In addition, for FDD channel prediction tasks, extrapolation is achieved in both the time and frequency domain, further increasing the difficulty.

{Unlike the existing approaches that designed entire networks specifically for channel prediction tasks,} in this paper, we attempt to adapt LLM for MISO-OFDM channel prediction to achieve improved predictive capability and generalization ability.
Specifically, we build a channel prediction neural network based on pre-trained GPT-2 and fine-tune it to predict the future downlink CSI sequence based on the historical uplink CSI sequence.
{
Unlike existing studies where LLMs are used for time series prediction, we fully consider the specific characteristics of the channel, and design preprocessor, embedding, and output modules to bridge the gap between CSI data and the LLM.}
In detail, considering multi-path effects, we process CSI from both the frequency and delay domain to extract the underlying physical propagation feature.
To fully preserve the general knowledge in the pre-trained LLM, most of its parameters are frozen during training.
Simulations evaluate the proposed method for both the TDD and FDD channel prediction tasks and demonstrate its superiority among existing baselines. 
The main contributions of our work are summarized as follows: 
\begin{itemize}
\item We propose a novel LLM-empowered channel prediction method (LLM4CP) for MISO-OFDM systems, i.e., fine-tuning pre-trained GPT-2 on channel prediction datasets.
To the best of our knowledge, it is the first attempt to adapt pre-trained LLM for channel prediction.
\item {In light of the unique channel characteristics, we design dedicated modules and processing to bridge the gap between channel data and the feature space of the LLM, thereby facilitating cross-modality knowledge transfer.} 
\item Preliminary results validate the SOTA performance of the proposed method on TDD/FDD channel prediction tasks. 
In addition, it demonstrates superior few-shot and generalization prediction performance, as well as low training and inference costs.

\end{itemize}

Notation: $(\cdot)^{\rm T}$, $(\cdot)^{\rm H}$, $(\cdot)^\dagger$, $\Vert\cdot\Vert$, and $\Vert\cdot\Vert_F$ denote the transpose, the conjugate transpose, pseudo-inverse, $l_2$ norm, and Frobenius norm, respectively. 
$\otimes$ represents the Kronecker product operation.
$\bm{a}[i]$ is the $i\mbox{-}$th element of a vector $\bm{a}$ and $\bm{A}[i,j]$ denotes the element of a matrix or tensor $\bm{A}$ at the $i\mbox{-}$th row and the $j\mbox{-}$th column. 
$\bm{A}[:,i]$ denotes the $i\mbox{-}$th column of $\bm{A}$.
\section{System Model}
We focus on a single-cell MISO-OFDM system comprising a base station (BS) and a mobile user.
The BS is equipped with a dual-polarized uniform planar array (UPA) with $N_{\rm t}=N_{\rm h}\times N_{\rm v}$ antennas, where $N_{\rm h}$ and $N_{\rm v}$ represent the number of antennas in the horizontal and vertical directions, respectively.
The user is equipped with an omnidirectional antenna and it can be readily applied for cases with multiple antennas through parallel processing.
The system is capable of operating in both TDD and FDD modes.
\subsection{Channel Model}
\begin{figure}[t]
	\center{\includegraphics[width=8.5cm]  {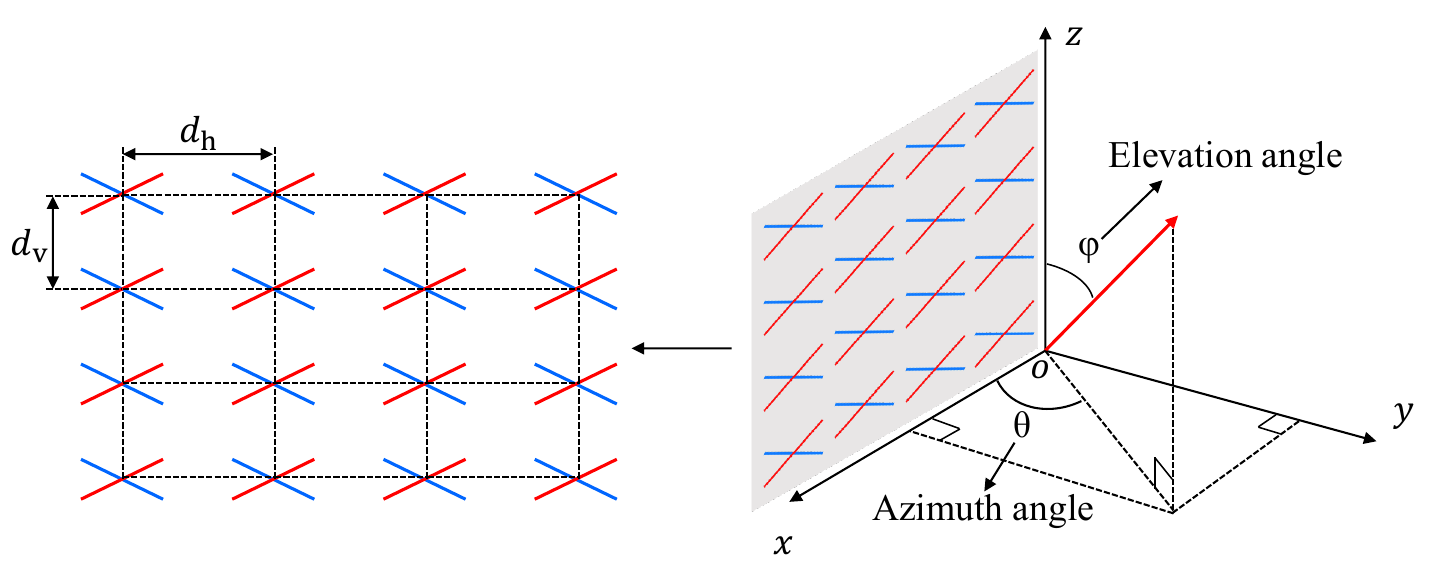}} 
	\caption{An illustration of the architecture of a dual-polarized UPA\upcite{yin2020addressing}.}
	\label{UPA}
\end{figure}
For a given polarization, adopting the classical cluster-based multi-path channel model\upcite{3gpp2018study}, the downlink CSI between BS and the user at time $t$ and frequency $f$ is
\begin{align}\label{CSI}
    \bm{h}(t,f)=\sum_{n=1}^N \sum_{m=1}^{M_n}\beta_{n,m}e^{j[2\pi (\upsilon_{n,m} t-f\tau_{n,m})+\Phi_{n,m}]}\bm{a}(\theta_{n,m},\varphi_{n,m}),
\end{align}
where $N$ and $M_n$ are the number of clusters and paths in each cluster, respectively.
$\beta_{n,m}$, $\upsilon_{n,m}$, $\tau_{n,m}$, and $\Phi_{n,m}$ represent the complex path gain, Doppler frequency shift, delay, and random phase of the $m\mbox{-}$th path of the $n\mbox{-}$th cluster, respectively.
{
Assuming the instantaneous speed of the mobile user at time $t$ is $v$ and the angle between the direction of the velocity vector and the path is $\phi_{m,n}$.
The Doppler frequency shift is derived as $\upsilon_{n,m}=\frac{vf\cos \phi_{m,n}}{2\pi c}$, where $c$ represents the velocity of light.
It is worth noting that the Doppler frequency shift caused by the user's movement is the main factor contributing to the time variation of the channel.
} 
$\bm{a}(\theta_{n,m},\varphi_{n,m})$ represents the steering vector of the corresponding path, where $\theta_{n,m}$ and $\varphi_{n,m}$ denote the azimuth and elevation angles.
Considering the structural characteristics of UPA as shown in Fig. \ref{UPA}, $\bm{a}(\theta_{n,m},\varphi_{n,m})$ is derived as
\begin{align}
\bm{a}(\theta_{n,m},\varphi_{n,m})=\bm{a}_{\rm h}(\theta_{n,m},\varphi_{n,m})\otimes\bm{a}_{\rm v}(\theta_{n,m}),
\end{align}
where $\bm{a}_{\rm h}(\theta,\varphi)$ and $\bm{a}_{\rm v}(\theta)$ are the horizontal and vertical steering vector, respectively.
$\bm{a}_{\rm h}(\theta,\varphi)[i]=e^{\frac{j2\pi i fd_{\rm h}\sin (\varphi)\cos (\theta)}{c}}$ and $\bm{a}_{\rm v}(\theta)[i]=e^{\frac{j2\pi i fd_{\rm v}\sin (\theta)}{c}}$, where $d_{\rm h}$ with $d_{\rm v}$ being the antenna spacing along the horizontal and vertical directions, respectively.
\subsection{Signal Model}
We consider a downlink MISO-OFDM signal transmission process, where $K_{\rm s}$ subcarriers are activated, with the $k\mbox{-}$th subcarrier denoted as $f_k$.
According to Eq. (\ref{CSI}), the downlink CSI at time $t$ and the $k\mbox{-}$th subcarrier is $\bm{h}_k=\bm{h}(t,f_k)$, which can be obtained through channel estimation or prediction.
Considering transmit precoding at the BS side, the received downlink signal of the $k\mbox{-}$th subcarrier at the user side is derived as
\begin{align}
y_k=\bm{h}_k^{\rm H}\bm{w}_k x_k + n_k,
\end{align}
where $n_k$ is the additive white Gaussian noise (AWGN) {with noise power $\sigma_n^2$} and $\bm{w}_k$ is the transmit precoder.
The achievable SE\upcite{marzetta2016fundamentals} of the downlink transmission process is derived as
\begin{align}\label{SE}
R=\sum_{k=1}^{K_{\rm s}}\log_2{\left(1+\frac{\lvert\bm{h}_k^{\rm H}\bm{w}_k\rvert^2}{\sigma_n^2}\right)}.
\end{align}
To maximize the downlink transmission rate, the matched-filtering based precoding is applied as
\begin{align}\label{ZF}
\bm{w}_k=\frac{\bm{h}_k}{\Vert\bm{h}_k\Vert}.
\end{align}
It is worth noting that inaccurate $\bm{h}_k$ will lead to mismatched $\bm{w}_k$, thereby impairing the SE.

\section{Problem Formulation for Channel Prediction}
In this section, we introduce a channel prediction-based transmission scheme and formulate a problem of inferring future downlink CSI based on historical uplink CSI.
\subsection{Channel Prediction-based Transmission}
\begin{figure}[t]
\center{\includegraphics[width=8.5cm]  {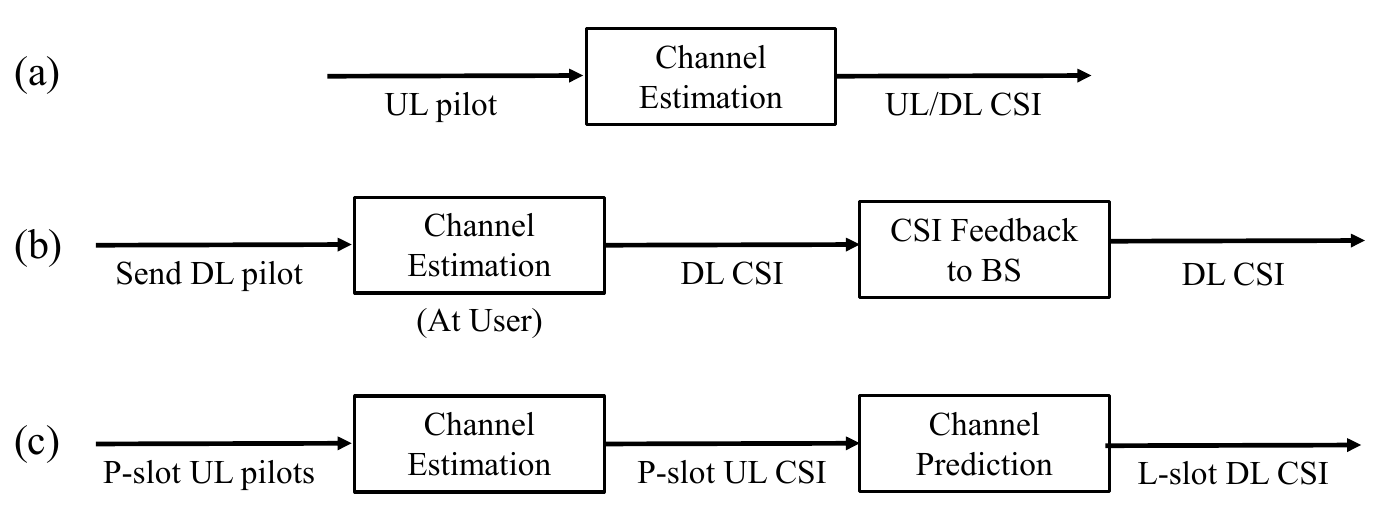}} 
\caption{An illustration of three schemes for downlink CSI acquisition. (a) Traditional downlink CSI acquisition scheme for TDD systems; (b) Traditional downlink CSI acquisition scheme for FDD systems; (c) Channel prediction-based downlink CSI acquisition for TDD/FDD systems.}
 \label{scheme}
\end{figure}
\begin{figure}[t]
\center{\includegraphics[width=8cm]  {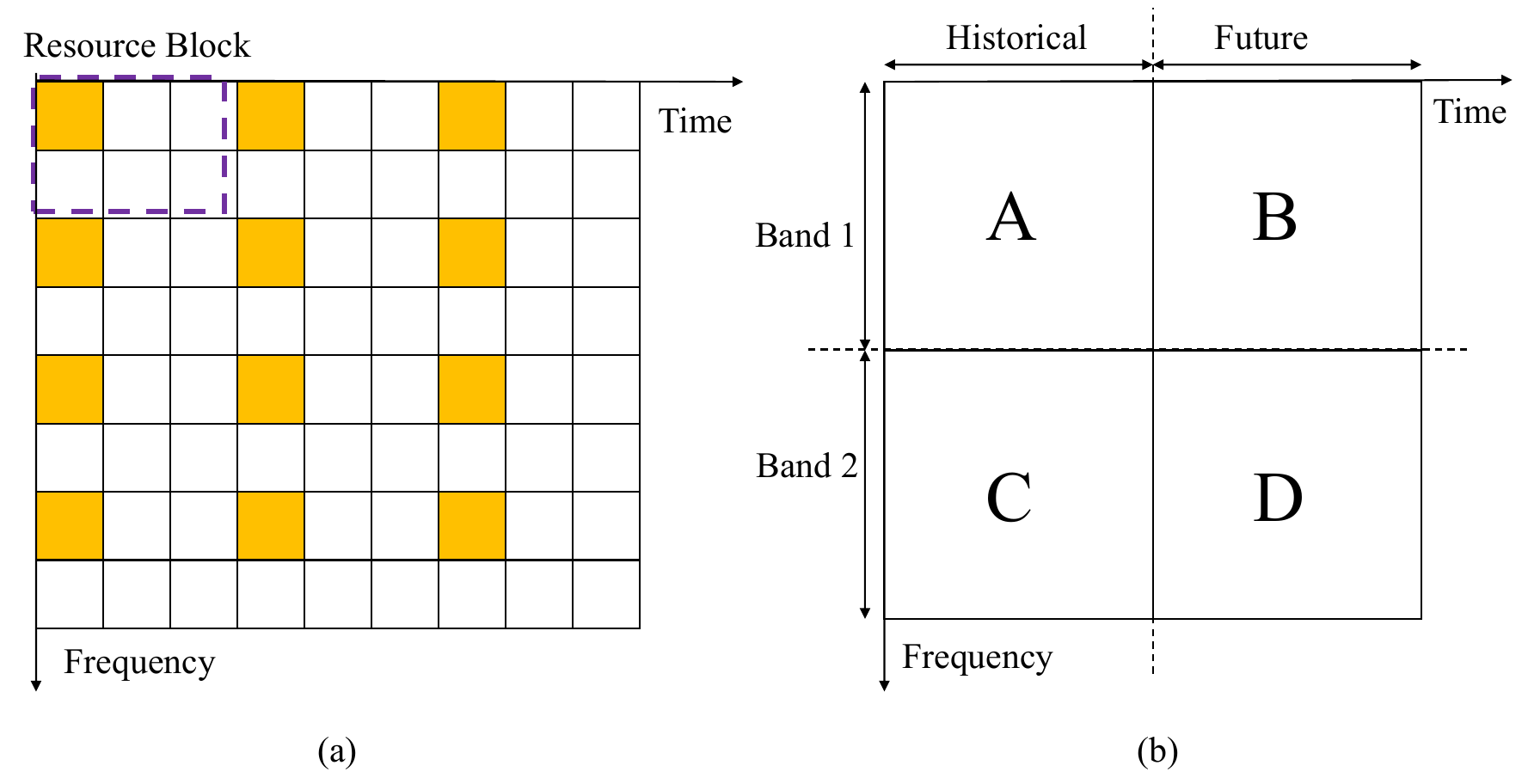}} 
\caption{(a) Resource block (RB); (b) An illustration of channel prediction in the time-frequency domain.}
 \label{RB}
\end{figure}
\begin{figure*}[t]
\center{\includegraphics[width=16cm]  {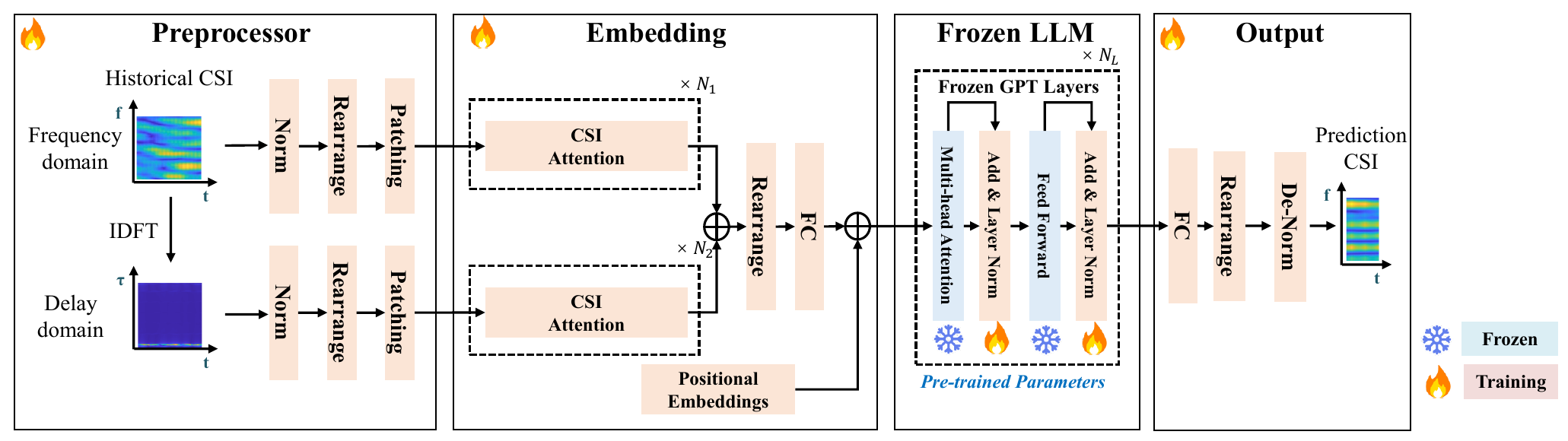}} 
\caption{An illustration of the network architecture of LLM4CP.}
 \label{network}
 \vspace{-5mm}
\end{figure*}
Traditional downlink CSI acquisition schemes for TDD and FDD systems are illustrated in Fig. \ref{scheme} (a) and (b), respectively.
In TDD systems, thanks to channel reciprocity, the downlink CSI can be obtained at the BS side by channel estimation on uplink pilots.
In FDD systems where the frequency of the uplink and downlink channels differs, downlink CSI can only be estimated at the user side and then fed back to the BS.
However, there are some shortcomings in existing downlink CSI acquisition methods. 
First, the CSI estimation and feedback process incur additional computational and transmission time overhead, causing channel aging\upcite{jiang2022accurate} in high dynamic scenarios.
In addition, extra downlink pilots occupy some of the time-frequency resources, reducing the SE of FDD systems.

Channel prediction-based transmission scheme provides a promising solution to address the above two drawbacks, as shown in Fig. \ref{scheme} (c).
Specifically, it predicts future downlink CSI sequences based on historical uplink CSI sequences,  avoiding the overhead of downlink pilots and feedback delay.
For further clarification, the time and frequency relationship between uplink and downlink CSI of the channel prediction-based scheme can be illustrated in Fig. \ref{RB} (b).
Region A represents the uplink CSI, while regions B and D correspond to the predicted downlink CSI under TDD and FDD modes, respectively. 
Each time-frequency region consists of multiple time-frequency resource blocks (RBs), and each RB contains a pilot, as shown in Fig. \ref{RB} (a).
In the following channel prediction process, we only consider the CSI associated with the pilots' positions, while CSI between pilots can be obtained through interpolation methods.
We assume that the uplink and downlink links have the same bandwidth and each covers $K$ resource blocks in the frequency domain.
In the time domain, future $L$ RBs are predicted based on historical $P$ RBs.
For simplicity, we denote the uplink and downlink CSI of each RB as $\bm{h}_{\rm u}^{k,s}$ and $\bm{h}_{\rm d}^{k,s}$, where $k$ and $s$ represent the indices of RBs in the frequency domain and time domain, respectively.
\subsection{Problem Formulation}
We aim to accurately predict future downlink CSI of $K \times L$ RBs based on historical CSI of $K \times P$ RBs.
The uplink CSI of $K$ subcarriers at time $i$ is represented in matrix form as
\begin{align}
\bm{H}_{\rm u}^{i}=[\bm{h}_{\rm u}^{1,i},...,\bm{h}_{\rm u}^{K,i}]^{\rm H}.
\end{align}
The downlink CSI can be obtained similarly as $\bm{H}_{\rm d}^{i}=[\bm{h}_{\rm d}^{1,i},...,\bm{h}_{\rm d}^{K,i}]^{\rm H}$.
The normalized MSE (NMSE)\upcite{jiang2022accurate} between predicted and actual downlink CSI is used to evaluate the prediction accuracy.
Then the entire problem can be described as  follows:
\begin{subequations}\label{problem}
\begin{align}
\min_{\Omega}\quad & {\rm NMSE}=\mathbb{E}\left\{\frac{\sum_{i=1}^L\Vert \hat{\bm{H}}_{\rm d}^{s+i}-\bm{H}_{\rm d}^{s+i} \Vert_F ^2}{\sum_{i=1}^L\Vert \bm{H}_{\rm d}^{s+i} \Vert_F ^2}\right\}\label{NMSE}\\
s.t.\quad&(\hat{\bm{H}}_{\rm d}^{s+1},...,\hat{\bm{H}}_{\rm d}^{s+L})=f_{\Omega}(\bm{H}_{\rm u}^{s},...,\bm{H}_{\rm u}^{s-P+1}),
\end{align}
\end{subequations}
where $\bm{H}_{\rm u}^{i}$, $\hat{\bm{H}}_{\rm d}^{i}$, and $\bm{H}_{\rm d}^{i}$ represent the estimated uplink CSI, predicted downlink CSI, and actual downlink CSI of RBs at time $i$, respectively.
$f_{\Omega}$ is the constructed mapping function and $\Omega$ is its variable parameters. 
In previous work, $f_{\Omega}$ represents either a parameterized model\upcite{truong2013effects, wong2005joint, shen2003short} or a deep learning network\upcite{jiang2022accurate,safari2019deep,zhang2023adversarial}. 
In this paper, we consider a novel pre-trained LLM-based channel prediction method to achieve higher prediction accuracy and generalization capability.
Specifically, $f_{\Omega}$ is the proposed LLM-based neural network and $\Omega$ represents the trainable parameters.
It is worth noting that the trained network is designed to handle the CSI prediction under two polarizations.

\section{LLM for Channel Prediction}
In this section, we propose an LLM-empowered MISO-OFDM channel prediction method, named LLM4CP, to predict the future downlink CSI sequence based on the historical uplink CSI sequence.
In order to adapt text-based pre-trained LLM to the complex matrix format of CSI data, specific modules are designed for format conversion and feature extraction, including preprocessor, embedding, backbone, and output.
The network components shown in Fig. \ref{network} and the training process are illustrated in detail below.

\subsection{Preprocessor Module}
Given the uplink CSI at time $i$ as $\bm{H}_{\rm u}^i\in \mathbb{C}^{K\times N_t}$, directly applying it to the network will bring significant complexity and training time, especially for a large number of antennas and subcarriers.
Therefore, we parallelize the processing of antennas, i.e., predicting the CSI for each pair of transmitter and receiver antennas separately.
For the $j\mbox{-}$th transmit antenna, the input sample of the network $\bm{H}_f\in \mathbb{C}^{K\times P}$ is derived as
\begin{align}
\bm{H}_f=[\bm{H}_{\rm u}^1[:,j],...,\bm{H}_{\rm u}^P[:,j]],
\end{align}
where $P$ is the temporal length of historical CSI.

Given that the delay domain, as referenced in Ref. \cite{yin2020addressing}, is the dual domain of the frequency domain, it serves to characterize the delay information of each multipath component, providing a more comprehensive perspective on frequency domain information\upcite{burghal2023enhanced}. Consequently, we simultaneously offer a representation in the delay domain through inverse discrete Fourier transform (IDFT) as well. 
\begin{align}
\bm{H}_{\tau}=\bm{F}_{K}^{\rm H}\bm{H}_f,
\end{align}
where $\bm{F}_K$ represents the $K$ dimensional DFT matrix. 
Since neural networks generally deal with real numbers, we convert $\bm{H}_f$ and $\bm{H}_{\tau}$ into real tensors $\bm{X}_f \in \mathbb{R}^{2\times K \times P}$ and $\bm{X}_{\tau}\in \mathbb{R}^{2\times K \times P}$, respectively.
To facilitate network training and convergence, we first normalize the input data as $\bar{\bm{X}_f}=\frac{\bm{X}_f-\mu_f}{\sigma_f}$ and $\bar{\bm{X}_{\tau}}=\frac{\bm{X}_{\tau}-\mu_{\tau}}{\sigma_{\tau}}$, where ($\mu_f$, $\mu_{\tau}$) and ($\sigma_{f}$, $\sigma_{\tau}$) represent the mean value and standard deviation of a batch of corresponding domain input data.
Then the tensors are rearranged to merge feature dimensions, i.e., $\tilde{\bm{X}_f}\in \mathbb{R}^{2K\times P}$ and $\tilde{\bm{X}_{\tau}}\in \mathbb{R}^{2K \times P}$.

\begin{figure}[t]
\center{\includegraphics[width=5cm]  {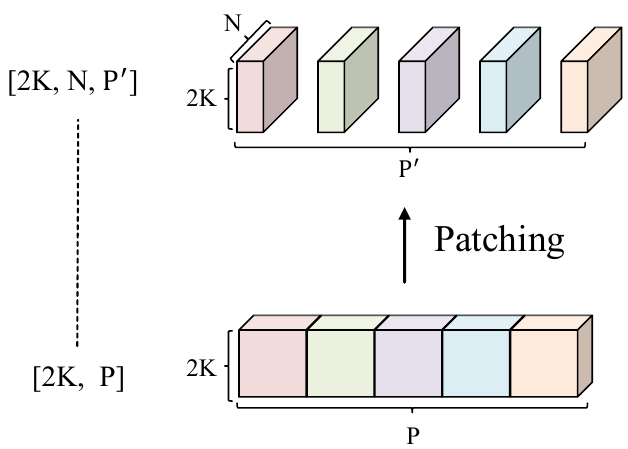}} 
\caption{An illustration of the patching operation.}
 \label{patching}
\end{figure}

To capture the local temporal features and reduce computational complexity, patching\upcite{nie2022time} operation along the temporal dimension is adopted, as shown in Fig. \ref{patching}.
Specifically, $\tilde{\bm{X}_f}$ and $\tilde{\bm{X}_{\tau}}$ are grouped into non-overlapping patches of size $N$ as the $\bm{X}_f^{p}\in \mathbb{R}^{2K \times N\times P'}$ and $\bm{X}_{\tau}^{p}\in \mathbb{R}^{2K \times N\times P'}$, where $P'=\lceil\frac{P}{N} \rceil$ is the number of patches.
Zero-padding is applied to the last patch that is not fully filled.
\subsection{Embedding Module}
\begin{figure}[t]
\center{\includegraphics[width=7cm]  {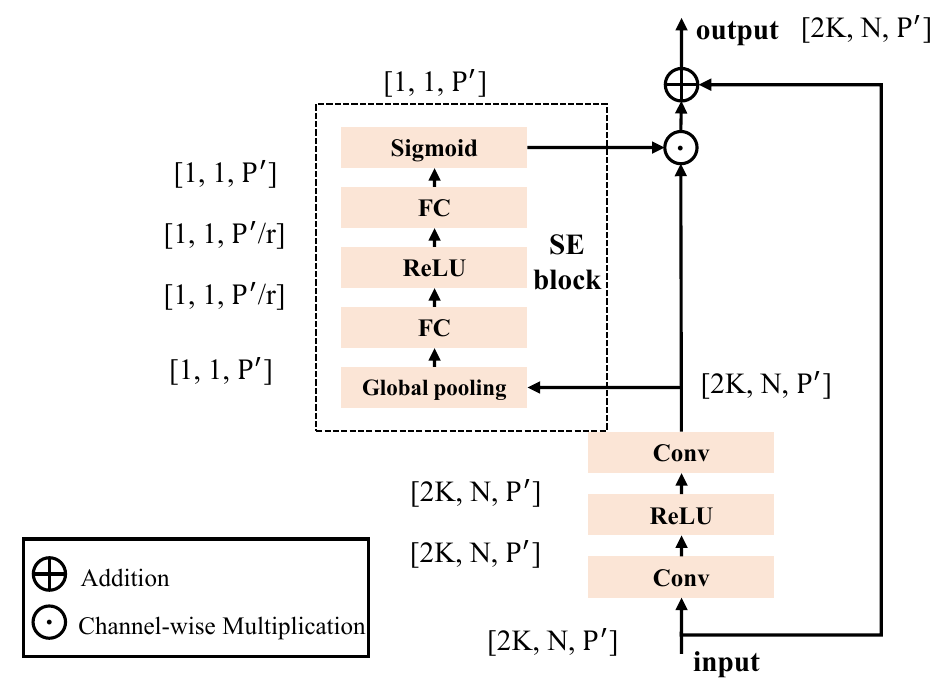}} 
\caption{{An illustration of the network structure and dimension changes of the CSI attention module. The network within the dashed box is the squeeze-and-excitation (SE) module.}}
 \label{TA}
\end{figure}
The embedding module is designed for preliminary feature extraction before  LLM, including CSI attention and position embedding.
First, the patched tensors are processed by corresponding CSI attention modules, as detailed below.

Inspired by Ref. \cite{hu2018squeeze}, the CSI attention module is designed for feature analysis, as shown in Fig. \ref{TA}.
{
The dimensions of each tensor during processing are described.}
For simplicity, we use $\bm{X}^i\in \mathbb{R}^{2K \times N\times P'}$ and $\bm{X}^o\in \mathbb{R}^{2K \times N\times P'}$ to represent the input and output tensors of the proposed CSI attention module, respectively.
First, the feature map is obtained as
\begin{align}
\bm{X}^{fm}={\rm Conv}({\rm ReLU}({\rm Conv}(\bm{X}^i)))\in \mathbb{R}^{2K \times N\times P'},
\end{align}
where $\rm Conv(\cdot)$ represents the 2D convolution operator and $\rm ReLU (\cdot)$ represents the ReLU\upcite{nair2010rectified} activation function.
The convolution layers extract temporal and frequency features within each patch and integrate features across different patches.
Then $\bm{X}^{fm}$ is passed through the SE block, which compromises squeeze and excitation parts, to obtain weights for different patches, i.e.,
\begin{align}
\bm{X}^{\rm SE}={\rm SE}(\bm{X}^{fm})\in \mathbb{R}^{1\times1\times P'}.
\end{align}
For the squeeze part, the global average pooling is first applied to generate
channel-wise statistics $\bm{X}^{\rm GAP}\in \mathbb{R}^{1\times1\times P'}$ and 
\begin{align}
\bm{X}^{\rm GAP}[i]=\frac{1}{2K \times P'}\sum_{j=1}^{2K}\sum_{k=1}^{P'}\bm{X}^{fm}[j,k,i].
\end{align}
For the excitation part, two fully connected (FC) layers are adopted to model the correlation between different patches.
To reduce computational complexity, the first FC layer shrinks the patch dimension size from $P'$ to $P'/r$ with $r\ge 1$, followed by a ReLU operation
Then, the second FC layer expands the patch dimension to $P'$.
Following this, the sigmoid function is utilized to generate the attention weight tensor $\bm{X}^{\rm SE}$, with each element falling within the range [0,1].
The scale operation weights each patch with $\bm{X}^{\rm SE}$ and obtains scaled features $\bm{X}^{\rm Sca}\in \mathbb{R}^{2K \times N\times P'}$, where
\begin{align}
\bm{X}^{\rm Sca}[:,:,i]=\bm{X}^{\rm SE}[i]\times\bm{X}^{fm}[:,:,i].
\end{align}
Then the residual connection is established as
\begin{align}
\bm{X}^{o}=\bm{X}^{\rm Sca}+\bm{X}^i.
\end{align}
It is worth noting that the CSI attention module can be concatenated multiple times to enhance feature extraction effectiveness.
The processed $\bm{X}_{f}^{p}$ and $\bm{X}_{\tau}^{p}$ are added as
\begin{align}
\bm{X}^{\rm CA}={\rm CA}^{(N_1)}(\bm{X}_{f}^{p})+{\rm CA}^{(N_2)}(\bm{X}_{\tau}^{p}),
\end{align}
where ${\rm CA}^{N}$ represents the CSI attention module cascaded $N$ times.
To adapt to the input format of LLM, $\bm{X}^{\rm CA}$ is first rearranged to $\tilde{\bm{X}}^{\rm CA}\in \mathbb{R}^{2KN\times P'}$ and then mapped to $\bar{\bm{X}}^{\rm CA}\in \mathbb{R}^{F\times P'}$ with a single FC layer, where $F$ is the feature dimension of the pre-trained LLM.
To incorporate positional information, non-learnable positional encoding\upcite{vaswani2017attention} $\bm{X}^{\rm PE}\in \mathbb{R}^{F\times P'}$ is introduced to enable the network to distinguish between patches at different positions.
Its structure is shown as
\begin{align}
\bm{X}^{\rm PE}(2i,j)=\sin \left(\frac{j}{10000^{\frac{2i}{F}}}\right),
\end{align}
and
\begin{align}
\bm{X}^{\rm PE}(2i+1,j)=\cos \left(\frac{j}{10000^{\frac{2i}{F}}}\right),
\end{align}
Finally, the embedding $\bm{X}^{\rm EB}\in \mathbb{R}^{F\times P'}$ is obtained as
\begin{align}
\bm{X}^{\rm EB}=\bar{\bm{X}}^{\rm CA}+\bm{X}^{\rm PE}.
\end{align}
\subsection{Backbone Network}
Recent studies\upcite{zhou2024one,ren2024tpllm} have revealed that pre-trained LLMs can be fine-tuned for cross-modal downstream tasks, including time series analysis.
In other words, during the pre-training process on extensive textual data, LLM has acquired general knowledge\upcite{zhou2024one}.
{
Inspired by this, we attempt to leverage the universal modeling capability of LLMs for channel prediction tasks.
However, the pre-trained LLM cannot directly process non-linguistic data due to the significant disparity between textual and CSI information.
Therefore, similar to how text is preprocessed into tokens by the tokenizer, we obtain the embeddings of CSI data using the proposed preprocessor and embedding module.
The preprocessed CSI ``tokens'' are then fed into the backbone of the LLM, i.e.,
\begin{align}
\bm{X}^{\rm LLM}={\rm LLM}(\bm{X}^{\rm EB})\in \mathbb{R}^{F\times P'},
\end{align}
where $\rm LLM(\cdot)$ denotes backbone networks of the LLM.}

Without loss of generality, GPT-2\upcite{radford2019language} is chosen as the LLM backbone in this work.
The backbone of GPT-2 is composed of a learnable positional embedding layer and stacked transformer decoders\upcite{vaswani2017attention},  where the number of stacks and feature dimensions can be flexibly adjusted according to the requirements.
Each layer consists of self-attention layers, feed-forward layers, addition, and layer normalization, as shown in Fig. \ref{network}.
During the training process, self-attention and feed-forward layers are frozen to retain universal knowledge, while addition, layer normalization, and positional embedding are fine-tuned for adapting the LLM to the channel prediction task.
It is worth noting that in the proposed method, the GPT-2 backbone can be flexibly replaced with other LLM, such as Llama\upcite{touvron2023llama}.
The selection of the type and size of the LLM needs to consider the trade-off between training costs and performance.

\subsection{Output Module}
The output is designed to convert the output features of the LLM into the final prediction results.
First, two FC layers are adopted to transform the dimension of $\bm{X}^{\rm LLM}$
\begin{align}
\hat{\bm{X}}={\rm FC}({\rm FC}(\bm{X}^{\rm LLM}))\in \mathbb{R}^{2K\times L},
\end{align}
where $\rm FC(\cdot)$ represents the FC layer and $L$ is the prediction length.
Then, $\hat{\bm{X}}$ is rearranged to $\bm{X}^{\rm re}\in \mathbb{R}^{2\times K\times L}$, where the first and the second dimension respectively correspond to the real part and the imaginary part.
Then $\bm{X}^{\rm re}$ is de-normalized to generate the final output of the network, i.e.,
\begin{align}
\bm{X}^{\rm de}=\sigma_f\bm{X}^{\rm re}+\mu_f,
\end{align}
and the final prediction result $\hat{\bm{H}}_f\in \mathbb{C}^{K\times L}$ is obtained as
\begin{align}
\hat{\bm{H}}_f=\bm{X}^{\rm de}[1,:,:]+j\bm{X}^{\rm de}[2,:,:],
\end{align}
where $j=\sqrt{-1}$ represents the imaginary unit. 
\subsection{Training Configuration}
The proposed neural network is initially trained on channel prediction datasets and then applied for testing.
In the training phase, the ground truth of the predicted CSI $\hat{\bm{H}}_f$ is available.
Then we obtain the ground truth of the network output as $\bm{X}^{gt}\in \mathbb{R}^{2\times K\times L}$.
The normalized mean square error (NMSE)\upcite{burghal2023enhanced} is adopted as the loss function to minimize the prediction error, i.e., 
\begin{align}
{\rm Loss}=\frac{\Vert\bm{X}^{\rm de}-\bm{X}^{\rm gt}\Vert_F^2}{\Vert\bm{X}^{\rm gt}\Vert_F^2}.
\end{align}
In addition, the validation loss also adopts the same loss function.
It is worth noting that the self-attention and feed-forward layers of the pre-trained GPT-2 are frozen, while the other parameters of the network are trainable, as shown in Figure \ref{network}.
Since the former includes the main parameters of the network, the number of trainable parameters is relatively small, which will be explained in detail in Section \ref{training cost}.
The model with the smallest validation loss is saved for the testing phase.
\section{Experiments}
In this section, we first illustrate the simulation settings and then evaluate the prediction performance of the proposed LLM4CP method.\footnote {The code is publicly available at {https://github.com/liuboxun/LLM4CP}}
\subsection{Simulation Setup}
\subsubsection{Dataset}
We adopt the widely used channel generator QuaDRiGa\upcite{jaeckel2014quadriga} to simulate time-varying CSI datasets compliant with 3GPP standards\upcite{3gpp2018study}.
We consider a MISO-OFDM system, where a BS is equipped with a dual-polarized UPA with $N_{\rm h}=N_{\rm v}=4$ and a user is equipped with a single omnidirectional antenna.
The antenna spacing is half of the wavelength at the center frequency.
We suppose the bandwidth of both the uplink and the downlink channels is 8.64 MHz and covers $K=48$ RBs, i.e., the frequency interval of pilots is 180 kHz.
For both the TDD and FDD modes, we set the center frequency of the uplink channel as 2.4 GHz.
For FDD modes, the uplink and downlink channels are adjacent.
We predict future $L=4$ RBs based on historical $P=16$ RBs and set the time interval of pilots as 0.5 ms.
We consider the 3GPP Urban Macro (UMa) channel model\upcite{3gpp2018study} and non-line-of-sight (NLOS) scenarios.
The number of clusters is 21 and the number of paths per cluster is 20.
The initial position of the user is randomized and the motion trajectory is set as linear type.
The training dataset and validation dataset respectively contain 8000 and 1000 samples, with user velocities uniformly distributed between 10 and 100 km/h.
The testing dataset contains 10 velocities ranging from 10 km/h to 100 km/h, with 1000 samples for each velocity.
\subsubsection{Baselines}
\begin{table}[t]
	\small
	\caption{Hyper-parameters for network training}
	\label{settings}
	\centering
	\begin{tabular}{c|c}
		\hline
		\textbf{Parameter} & \textbf{Value} \\ \hline
		Batch size  & 512 \\ \hline
		Epochs & 500 \\ \hline
		Optimizer & Adam (betas=(0.9, 0.999)) \\ \hline
		Starting learning rate & 0.001 \\ \hline
		Learning rate decay rate & 0.1 per 150 epochs \\ \hline
	\end{tabular}
\end{table}
To validate the superiority of the proposed method, several model-based and deep learning-based channel prediction methods are implemented as baselines.
\begin{itemize} 
\item \textbf{PAD}\upcite{yin2020addressing}: PAD is an advanced model-based channel prediction method to overcome the curse of mobility for TDD systems.
In this approach, the order of the predictor is set as $N=8$, and historical $2N-1=15$ CSI samples are utilized for parameter calculation\upcite{yin2020addressing}.
\item \textbf{RNN}\upcite{jiang2019neural}: RNN is a classical neural network for series processing and has been adopted for channel prediction tasks.
In experiments, the number of RNN layers is set as 4.
\item \textbf{LSTM}\upcite{jiang2020deep}: LSTM is designed with memory cells and multiplicative gates to deal with long-term dependency.
An LSTM-based predictor with 4 LSTM layers is applied. 
\item \textbf{GRU}\upcite{cho2014learning}: Gate recurrent unit (GRU)\upcite{cho2014learning} is a variant of LSTM to address the vanishing gradient problem.
Similarly, the number of GRU layers is set as 4 for channel prediction.
\item \textbf{CNN}\upcite{safari2019deep}: In Ref. \cite{safari2019deep}, a CNN-based predictor is proposed for FDD systems, treating the prediction process of time-frequency CSI data as a two-dimensional image processing task.
This CNN model consists of ten convolutional layers, with a convolution kernel size of $3\times 3$.
\item \textbf{Transformer}\upcite{jiang2022accurate}: In Ref. \cite{jiang2022accurate}, a transformer-based parallel channel predictor is proposed for TDD systems to avoid error propagation problems.
Additionally, it is used as a basis for comparison.
\item \textbf{No prediction}: For the case of no prediction, we directly take the value of the future downlink $L$ CSI as the value of the latest uplink CSI to illustrate the severity of channel aging.
\end{itemize}

To ensure fairness, the above deep learning-based methods process antenna dimensions in parallel and adopt NMSE as the loss function for training.
Since deep learning-based approaches do not rely on prior assumptions, all schemes except for PAD are applied to both TDD and FDD modes.
\subsubsection{Network and Training Parameters}
In the simulation, we adopt $N_1=N_2=4$ CSI attention module for frequency and delay domain data processing.
For patching, the size of patches is set as $N=4$, and historical CSI is grouped into $P'=P/N=4$ non-overlapping patches.
For the SE block, the value of the reduction ratio is set as 2 and the convolution kernel size is set as $3\times 3$ for all convolution layers.
The smallest version\upcite{vaswani2017attention} of GPT-2 with $F=768$ feature dimension is adopted, the first $N_L=6$ layers of which are deployed.
Several hyper-parameters for model training are illustrated in Table \ref{settings}.
\subsubsection{Performance Metric}
In order to comprehensively evaluate the performance of the proposed scheme, we adopt three performance metrics, namely NMSE, SE, and bit error rate (BER).
\begin{itemize}
\item \textbf{NMSE}: 
As shown in Eq. (\ref{NMSE}), NMSE is a widely used\upcite{burghal2023enhanced} performance metric to directly characterize the accuracy of channel prediction.
Therefore, NMSE is adopted as a major metric in our experiments.
\item \textbf{SE}: SE is an important metric that reveals the achievable rate of the system, reflecting the effectiveness of communication.
It is calculated by Eq. (\ref{SE}), where $\bm{h}_k$ is the actual CSI and $\bm{w}_k$ is obtained as Eq. (\ref{ZF}) with predicted $\bm{h}_k$.
The communication SNR is defined as $1/{\sigma_n^2}$ and set as 10dB. 
\item \textbf{BER}: BER describes the communication reliability at a certain transmission rate.
During the simulation, 4-QAM (quadrature amplitude modulation) is adopted and the communication SNR is set as 10 dB.

\end{itemize}
\subsection{Performance Evaluation}
\begin{figure}[t]
\center{\includegraphics[width=7.5cm]  {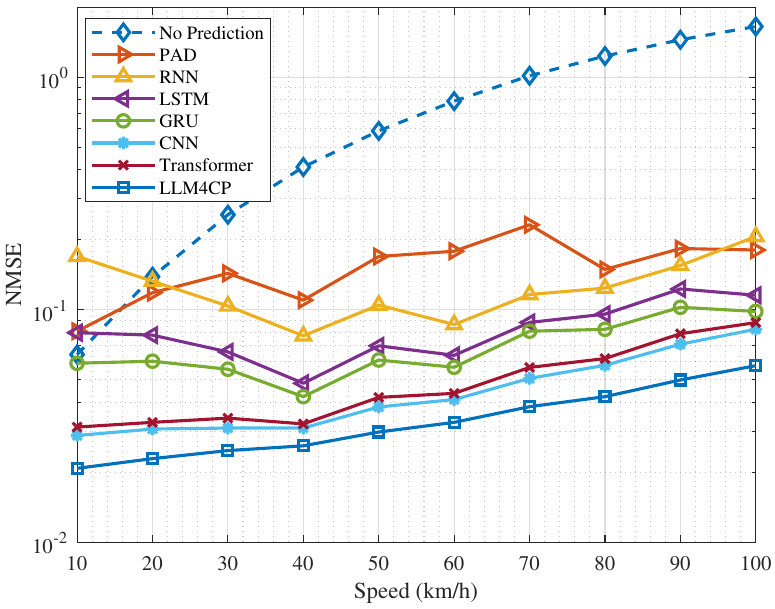}} 
\caption{The NMSE performance of LLM4CP and other baselines versus different user velocities for TDD systems.}
 \label{1_1}
\end{figure}
\begin{figure}[t]
\center{\includegraphics[width=7.5cm]  {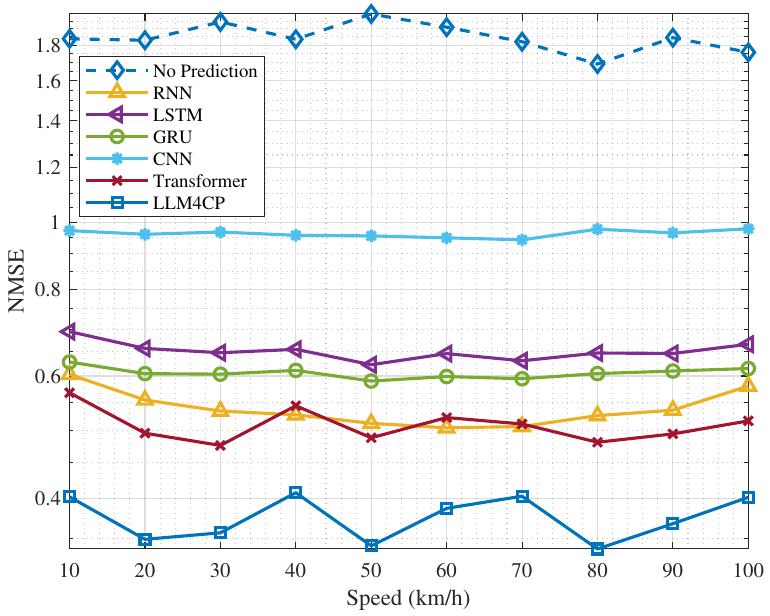}} 
\caption{The NMSE performance of LLM4CP and other baselines versus different user velocities for FDD systems.}
 \label{1_2}
\end{figure}

\begin{figure}[t]
\center{\includegraphics[width=7.5cm]  {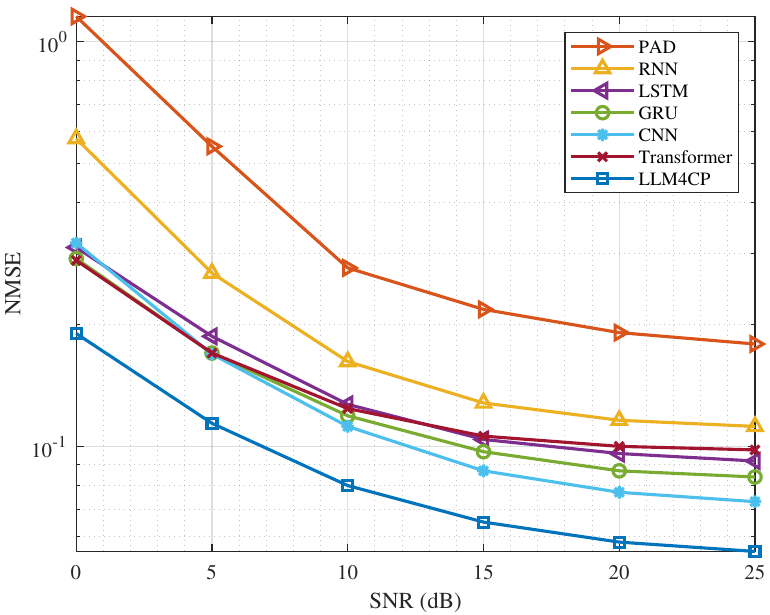}} 
\caption{The NMSE performance versus SNR of nosiy historical CSI for TDD systems.}
 \label{2_1}
\end{figure}

\begin{figure}[t]
\center{\includegraphics[width=7.5cm]  {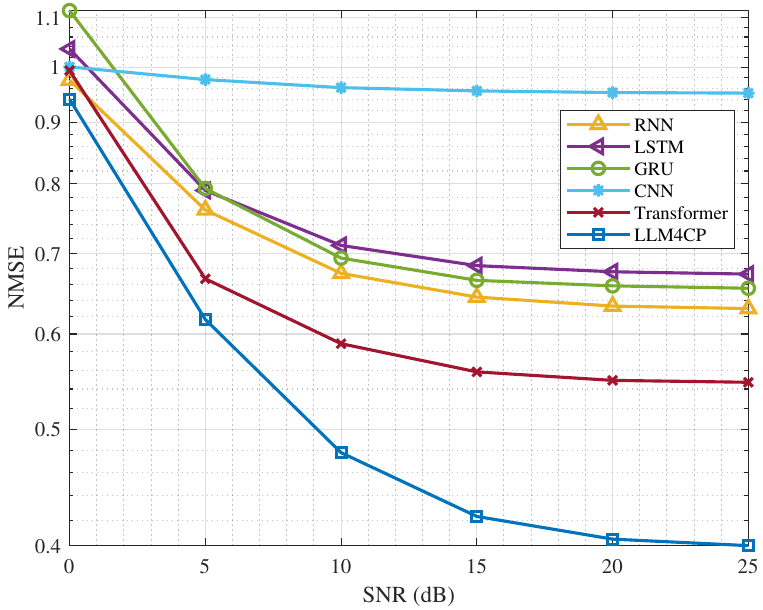}} 
\caption{The NMSE performance versus SNR of nosiy historical CSI for FDD systems.}
 \label{2_2}
\end{figure}

\begin{table*}[t]
\small
\caption{The SE and BER performance of LLM4CP and other baselines for TDD systems. (Maximum achievable SE: 7.33 bps/Hz)}
\label{1_1_table}
\centering
\begin{tabular}{|c|c|c|c|c|c|c|c|c|}
\hline
Metric & No Prediction & PAD & RNN & LSTM & GRU & CNN & Transformer & LLM4CP \\ \hline
SE (bps/Hz) & 6.238  &  7.007 &6.692 & 6.816
&6.772 & 6.992 & 6.963  & \textbf{7.036} \\ \hline
BER & 0.1259 & 0.0050 & 0.0208& 0.0148 & 0.0189 & 0.0065 & 0.0081 & \textbf{0.0039} \\ \hline
\end{tabular}
\end{table*}
\begin{table*}[t]
\small
\caption{The SE and BER performance of LLM4CP and other baselines for FDD systems. (Maximum achievable SE: 7.33 bps/Hz)}
\label{1_2_table}
\centering
\begin{tabular}{|c|c|c|c|c|c|c|c|}
\hline
Metric & No Prediction & RNN & LSTM & GRU & CNN & Transformer & LLM4CP \\ \hline
SE (bps/Hz) & 4.992 & 5.336 & 5.306&5.292 & 4.119 & 5.754  & \textbf{6.303} \\ \hline
BER & 0.3189 & 0.1313 & 0.1303 & 0.1315 & 0.2832 & 0.0807 & \textbf{0.0347} \\ \hline
\end{tabular}
\end{table*}
\begin{table*}[ht]
	\small
	\centering
	\caption{Results of ablation experiments.}
	\label{ablation}
	\begin{tabular}{|c|c|c|c|c|c|}
		\hline
		& LLM4CP & w/o delay domain & w/o CSI attention & w/o patching & w/o LLM \\ \hline
		NMSE & \textbf{0.043} & 0.045 & 0.053 & \textbf{0.043} & 0.049 \\ \hline
		SE (bps/Hz) & \textbf{7.073} & 7.051 & 7.036 & 7.066 & 7.036 \\ \hline
		BER & \textbf{0.0031} & 0.0036 & 0.0049 & 0.0036 & 0.0048 \\ \hline
	\end{tabular}
\end{table*}
\begin{table*}[t]
	\small
	\caption{Network parameters (training parameters/total parameters) and the training/interference cost per batch.}
	\label{cost}
	\centering
	\begin{tabular}{|c|c|c|c|c|c|c|c|}
		\hline
		& PAD & RNN & LSTM & GRU & CNN & Transformer & LLM4CP \\ \hline
		Network parameters (M) &  0/0 & 0.30/0.30 & 1.13/1.13 &0.86/0.86 & 3.14/3.14 & 1.76/1.76  & 1.73/82.87 \\ \hline
		Training time (ms) & 0 & 4.59 & 7.40 & 5.88 & 2.03
		& 21.84 & 6.84 \\ \hline
		Interference time (ms) & 31.96 & 3.55& 5.19 & 4.02 & 0.52 & 18.35 & 5.84 \\ \hline
	\end{tabular}
\end{table*}
\begin{figure}[t]
	\center{\includegraphics[width=7.5cm]  {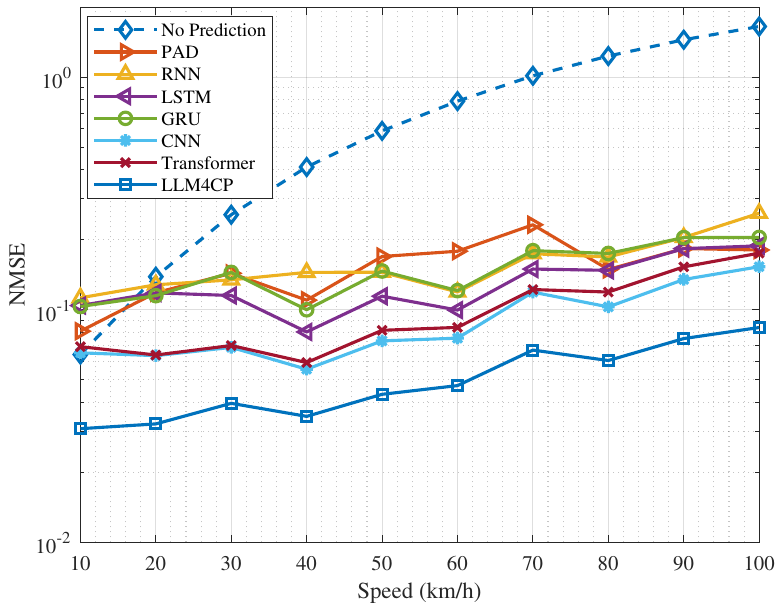}} 
	\caption{The NMSE performance of LLM4CP and other baselines versus different user velocities for TDD systems (Few-shot: 10\% training dataset).}
	\label{3_1}
\end{figure}
\begin{figure}[t]
	\center{\includegraphics[width=7.5cm]  {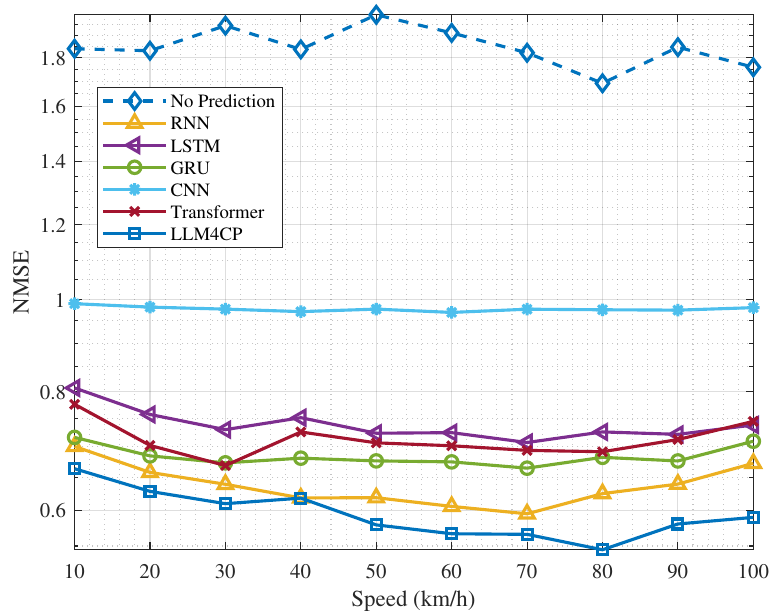}} 
	\caption{The NMSE performance of LLM4CP and other baselines versus different user velocities for FDD systems (Few-shot: 10\% training dataset).}
	\label{3_2}
\end{figure}
\begin{figure}[t]
\center{\includegraphics[width=7.5cm]  {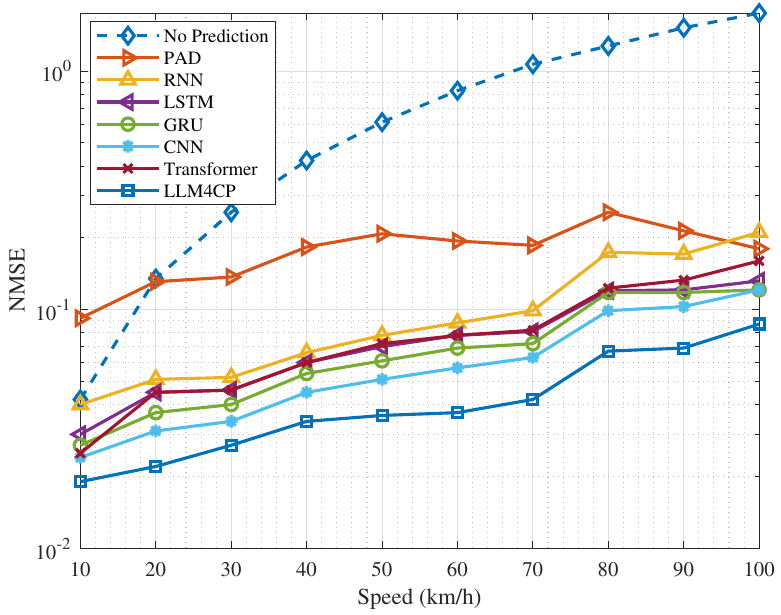}} 
\caption{The zero-shot generalization performance for the TDD systems in UMi scenario.}
 \label{4_1}
\end{figure}
\begin{figure}[t]
\center{\includegraphics[width=7.5cm]  {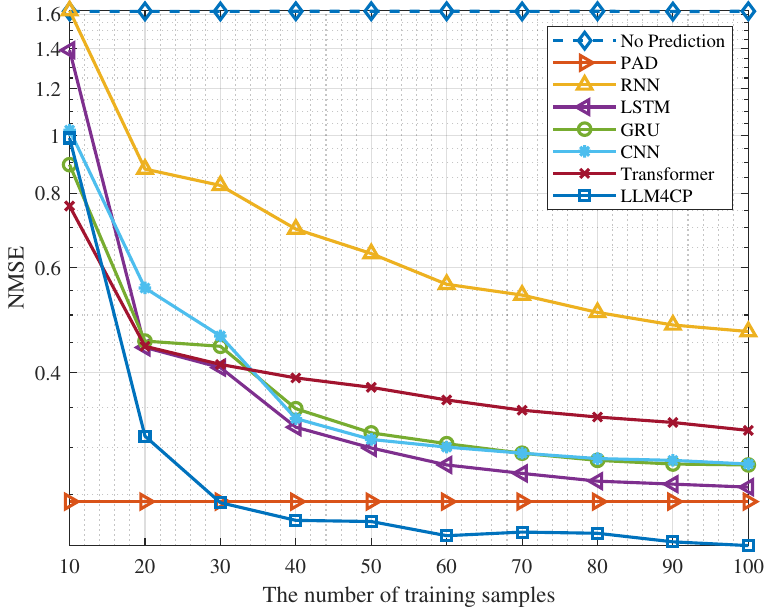}} 
\caption{The cross-frequency generalization performance versus different training samples for the TDD systems.}
 \label{4_2}
\end{figure}
\subsubsection{Performance Under Varying Velocities}
For TDD systems, we compare the NMSE performance of the proposed LLM4CP with baselines among different user velocities, as shown in Fig. \ref{1_1}.
It can be observed that as the user velocity increases, the NMSE of all baselines gradually increases.
This is due to that as the user velocity increases, the CSI changes rapidly with a shorter channel coherence time, resulting in increased prediction difficulty.
Directly adopting the latest CSI as the predicted CSI will bring huge prediction errors, especially for highly dynamic scenes.
The proposed method consistently outperforms other baselines among testing velocities and demonstrates its high prediction accuracy.

For FDD systems, the NMSE performance among different user velocities is given in Fig. \ref{1_2}.
Due to the frequency gap between the uplink and downlink channels, the no-prediction scheme is unacceptable for the FDD mode.
LLM4CP shows obvious advantages under all  user velocities.
It is worth noting that compared with TDD systems, the superiority of LLM4CP is greater in FDD systems, thanks to its powerful ability to model complex time-frequency relationships.

In addition, to evaluate the communication effectiveness and reliability of these methods, both the SE and BER performance are presented for TDD and FDD systems, as shown in Table. \ref{1_1_table} and \ref{1_2_table}, respectively.
The maximum achievable SE is obtained with perfect CSI.
Both the SE and BER have been averaged over all test speeds.
Compared with TDD mode, the performance of all baselines in FDD mode shows an overall decrease due to the greater difficulty of prediction.
Nevertheless, the proposed method achieves SOTA SE and BER performance for both the TDD and FDD modes.

\subsubsection{Robustness Against Noise}
Due to the inaccuracy of channel estimation, the predictor's robustness against noise is crucial, that is, predicting future CSI based on noisy historical CSI.
In Fig. \ref{2_1} and \ref{2_2}, the NMSE of these methods with noisy historical CSI is provided for both the TDD and FDD systems.
Specifically, in the testing phase, the historical CSI data is added by Gaussian white noise with variance $\sigma_{\rm n}^2$ and SNR is defined as $1/{\sigma_{\rm n}^2}$.
To enhance the robustness, during the training phase, noise with SNR uniformly distributed between 0 and 25dB is added to all baselines.
It can be observed that for all schemes, lower SNR results in higher prediction NMSE.
The proposed method exhibits the lowest NMSE at most SNRs, indicating its high robustness performance against CSI noise.

\subsubsection{Few-shot Prediction}
To reduce the cost of CSI data collection and network training, few-shot prediction is crucial for the rapid deployment of deep learning-based models.
In this part, we evaluate the few-shot learning ability of the proposed methods, where only 10\% of the dataset is utilized for network training.

In Fig.\ref{3_1}, the few-shot performance of the proposed LLM4CP and other baselines for TDD systems is illustrated.
The powerful few-shot learning capability of LLM enables the proposed method to perform well even with limited training samples. 
When compared to full-sample training, as demonstrated in Fig. \ref{1_1}, the advantages of LLM4CP over other baselines are more evident in few-shot prediction scenarios.
In Fig. \ref{3_2}, the few-shot performance for FDD systems is also given. 
Even though FDD channel prediction is more difficult, the proposed method maintains its advantages at most speeds.

\subsubsection{Generalization Experiments}
Generalization ability is crucial for model deployment in real-world scenarios, where well-trained models are applied to new scenarios with few-shot or even zero-shot training.
In Fig. \ref{4_1}, considering TDD systems, we directly apply the model trained in the Uma scenario to the 3GPP Urban Micro (UMi) scenario\upcite{3gpp2018study} without additional training process, while keeping the other settings unchanged.
The proposed model surpasses other baselines in terms of NMSE metric, demonstrating its strong generalization capability across different channel distributions.
Furthermore, we consider the model's cross-frequency generalization capability, which is important for practical multi-band communication systems. 
Specifically, we apply the model trained at 2.4 GHz TDD systems to 4.9 GHz TDD systems with few-shot and zero-shot training.
In Fig. \ref{4_2}, all deep learning-based baselines show poor zero-shot learning capabilities due to the channel's striking frequency selectivity.
Nevertheless, as the number of training samples increases, the prediction NMSE of the proposed scheme drops significantly and surpasses other schemes.
The proposed LLM4CP method achieves great NMSE performance at the new frequency with only a few training samples, demonstrating its strong frequency generalization ability.

\subsubsection{Ablation Experiments}

\begin{table*}[t]
	\small
	\caption{{The NMSE performance, network parameters, and interference time of LLM4CP with different numbers of GPT-2 layers. (LLM4CP ($n$) represents the proposed method with $n$ GPT-2 layers.)}}
	\label{number layers}
	\centering
	\begin{tabular}{|c|c|c|c|}
		\hline
		 & NMSE & Network parameters (M) & Interference time (ms) \\ \hline
		LLM4CP (2) & 0.0626 & 1.72/54.49 & 3.67 \\ \hline
		LLM4CP (4) & 0.0568 & 1.73/68.66 & 5.02 \\ \hline
		LLM4CP (6) & 0.0509 & 1.73/82.87 & 5.81 \\ \hline
		LLM4CP (8) & 0.0610 & 1.74/97.01 & 6.77 \\ \hline
	\end{tabular}
\end{table*}
To validate the effectiveness of several specific modules, the ablation experiment is conducted by removing relevant modules, as shown in Table \ref{ablation}.
The NMSE, SE, and BER performance for TDD systems are given and averaged over all testing speeds.
For LLM4CP without delay domain processing, the subsequent network of the delay domain transformation is removed.
For LLM4CP without CSI attention modules, the patched tensors of both the frequency and the delay domain are directly added for the following processing.
For LLM4CP without patching, the output of the preprocessor is directly input into CSI attention modules.
For LLM4CP without LLM, the frozen LLM is removed while other parts remain unchanged.
Observably, the removal of any of these four modules results in a loss of performance, indicating the necessity of these modules for high predictive accuracy.
\subsubsection{Training and Inference Cost}\label{training cost}
We compare the model training and inference cost of the proposed method with other baselines to assess the difficulty of deploying the model in practical scenarios, as shown in Table \ref{cost}.
All experiments are conducted on the same machine with 4 Intel Xeon Platinum 8358P CPUs, 4 NVIDIA GeForce RTX4090 GPUs, and 188 GB of RAM.
Since PAD is a model-based method, the number of its model parameters is relatively small to neglect and the training process is not required.
However, it takes the longest inference time due to the high processing complexity.
Although LLM4CP has large total parameters, the actual trainable parameters are similar to those of other deep learning-based models since most parameters in the LLM are frozen.
It is worth noting that LLM's inference time is much shorter than that of the Transformer thanks to inference acceleration specific to the GPT model.
Therefore, the proposed LLM4CP has the potential to serve real-time channel prediction.

{
In addition, we comprehensively evaluated the impact of selecting different numbers of GPT-2 layers on channel prediction performance, parameters cost, and inference time, as shown in Table \ref{number layers}.
The relevant experimental results are obtained under the TDD few-shot prediction settings with 10\% training dataset, where NMSE is the average across different speeds.
The network parameters and inference time are both increased with the number of GPT-2 layers.
It is worth noting that the proposed model with 6 GPT-2 layers performs the best within the testing range, indicating that having more layers does not necessarily favor prediction.
In practical deployment, the selection of the type and size of the LLM backbone needs to consider both the requirements for prediction accuracy and the constraints of device storage and computational resources.
}

\section{Conclusions and Future Work}
In this paper, we have proposed an LLM-empowered channel prediction method, which fine-tunes pre-trained GPT-2 for MISO-OFDM channel prediction tasks.
It predicts the future downlink CSI sequence based on the historical uplink CSI sequence and can be applied for both the TDD and FDD systems.
To account for channel characteristics, we have tailored preprocessor, embedding, and output modules to bridge the gap between CSI data and LLM, with the aim of fully leveraging the transfer knowledge across models from the pre-trained LLM. 
Preliminary simulations validate its superiority over existing model-based and deep learning-based channel prediction methods in full-sample, few-shot, and generalization tests with acceptable training and inference costs.

In the future, we plan to explore more comprehensive experimental setups and validate the proposed method using a more realistic and challenging CSI dataset. Additionally, we will incorporate link-level simulation with channel coding and evaluate the frame error rate (FER) of the system.

\end{document}